# Short spatial mooring-tilt variations from deep Mediterranean observations

**by Hans van Haren**

Royal Netherlands Institute for Sea Research (NIOZ), P.O. Box 59, 1790 AB Den Burg, the Netherlands.
e-mail: hans.van.haren@nioz.nl

**Abstract.** Interaction between energy-abundant mesoscale eddies and internal waves can lead to convection-turbulence generation and may prove important for deep-sea life and circulation. However, the size of scales of interacting flows is not well known. In this paper, a diagnostic tool of tilt is tested near the single top-buoyancy of 40 mooring lines 9.5 m apart horizontally and compared with 50-m scale relative vorticity and waterflow above a 2500-m deep flat Northwestern-Mediterranean seafloor. Whilst tilt relates to first order with flow-speed squared induced by mooring-line drag, considerable deviations from this relationship and larger tilt occur when the amplitude of relative vorticity attains values O(f), f the inertial frequency of planetary vorticity. The sign of relative vorticity is of no importance. Such larger-tilt events occur during most intense convection turbulence via warm-water slanting from above. During these events, variations in tilt-angle magnitude are as large as the average tilt O(0.1)°, thereby reducing variational scales from 50 to 9.5 m. Thus, in a deep-sea environment where flow speeds are <0.07 m $s^{-1}$, O(0.01) m $s^{-1}$ flow-speed variations provide important turbulent mixing, under conditions without deep dense-water formation.

**Keywords:** convection turbulence; deep Mediterranean Sea; mooring-line tilt observations; short spatial scales

## 1 Introduction

In a stably stratified environment like the sun-heated ocean, downward pulses of warm water seem impossible in terms of irreversible turbulent convection. In the deep weakly stratified Northwestern Mediterranean Sea however, such pulses are observed to dominate local turbulent mixing about 40% of time, governed by sub-mesoscale eddies [1] and inertio-gravity waves IGW [2], underneath a large-scale boundary flow that varies at mesoscales [3]. The convection-type turbulence is found on average three times more intense than geothermal heating through the seafloor, and not associated with deep dense-water formation that was absent [2].

The region also shows extensive generation of near-inertial waves, which under conditions of weakly stratified waters can lead, like eddies, to slantwise convection, as has been proposed in general [4], [5] and inferred from shipborne profiling observations in the open Western Mediterranean [6]. The convection may develop to irreversible 3D turbulence, while vertical (opposite-to-gravity) density profiles appear stably stratified [5], [7]. Under such conditions, the traditional internal-wave bounds [f, N] of inertial $f = 2\Omega\sin\varphi$ and buoyancy N frequencies extend to IGW bounds [$\omega_{min}$ $\omega_{max}$] [8]. Here, $\varphi$ denotes latitude, and $\Omega$ the Earth rotational frequency. For meridional wave propagation, the $\omega_{min}$<f and $\omega_{max}$>2$\Omega$ or >N, whichever is largest of N and 2$\Omega$, are functions of N, or shear magnitude $|S| \sim N$ in regularly observed marginal stability conditions, e.g., [9], of $\varphi$, and of direction of wave propagation [8], [10],

$$\omega_{max}, \omega_{min} = (A \pm (A^2 - B^2)^{1/2})^{1/2}/\surd 2, \quad (1)$$

in which $A = N^2 + f^2 + f_s^2$, $B = 2fN$, and $f_s = f_h\sin\alpha_\varphi$, $f_h = 2\Omega\cos\varphi$, and $\alpha_\varphi$ the angle to zonal direction. As follows from (1), inertial waves are the only IGW that can propagate in all N, all $\alpha_\varphi$.

In the deep Mediterranean N = O(f). The associated IGW band (1) thus extends free-propagating wave frequencies to sub-inertial, sub-mesoscales, for meridional propagation. In theory, the bounds may be further extended locally under horizontal flow differences generating the vertical component of relative vorticity, e.g., [11],

$$\zeta = \partial V/\partial x - \partial U/\partial y, \quad (2)$$

for horizontal waterflow components (U, V) in East and North directions, respectively. Depending on its sign, this relative vorticity may add to or subtract from planetary vorticity component f. In the Northwestern Mediterranean values of $|\zeta| = f/2$ are reported from mid-depth drifter observations [1]. The question is what its value amounts to in the deep, very weakly stratified Mediterranean and what its scales are compared to that of the motions it might trap or reflect.

In this paper, observational investigations of convective water pulses are further pursued that occur frequently in the deep Northwestern Mediterranean. An unusual tool of vertical mooring-line tilt is tested for its usefulness to study vibrations that may associate with convection turbulence. For this purpose, additional sensors were installed at two levels in a 3D temperature-sensor mooring array [12]. At the top and bottom of each of the closely spaced 45 mooring lines a tilt sensor was installed. Although typical tilt-angle variations are only ±0.1° to the vertical in waterflow with speeds < 0.07 m $s^{-1}$, these reflect large-scale dynamical variations that are driven by mesoscale processes. As will be demonstrated, between the lines at 9.5-m horizontal distance however, variations are similar as at large scales, which reflects the importance of small-scale variations.

## 2 Materials and Methods

A 'large-ring mooring' array was deployed at the <1° flat and 2458-m deep seafloor of 42° 49.50′N, 006° 11.78′E just 5 km south of the foot of the steep continental slope of the Northwestern Mediterranean Sea (Fig. 1a), in October 2020. Details of mooring layout and deployment are given in [12] and of data processing in [2]. In summary, 65 high-resolution NIOZ4 temperature sensors [13] were taped at 2-m intervals to each of 45 vertical lines of 0.006-m diameter and 125-m tall that were tensioned to 1.3 kN by a single 0.4-m-diameter cylinder buoy on top. The lines were attached at 9.5-m horizontal intervals to a steel-cable grid that was tensioned inside a 70-m diameter steel-tube ring (Fig. 1b), which functioned as a float prior to deployment and as an anchor after landing at the seafloor.

Every line held two T-sensors, one just below the buoy (Fig. 2) and another near the cable-grid, of which the tilt sensor was also activated. These T-tilt sensors effectively sampled at a rate of once per 4 s. This allowed for the monitoring of the mooring-line deviation from vertical at two levels, a lower at about h = 1.5 m above seafloor and an upper at about h = 125.5 m. Due to unknown causes all T-sensors switched off unintentionally when the file-size on the memory card reached 30 MB after 7500 startups for writing a 4-kB data block. It implied that a maximum of about three months of temperature-tilt data was obtained. (The temperature-only sensors recorded about 20 months of data.) In total 8 out of 90 tilt sensors failed for various electronic reasons. The analysis in this paper focuses on the upper-level T-sensor data, and their short variations with time and space. Besides time series and spectral analyses, 2D and quasi-3D movies are presented, for their construction see [2].

Although tilt is recorded along three axes to provide a vector, only its magnitude, more precisely the relative magnitude of its angle to the vertical $|\alpha|$[°], is considered here as a proxy for mooring vibrations and the various dynamic waterflow processes causing them. While this follows from practical reasons since a recent calibration is not available and electronic drift is about 0.1° per 3 months, the tilt-vector direction is of no use since the mooring line can rotate freely due to a swivel. The T-tilt data are referenced to current-meter tilt data obtained simultaneously at the upper level only (Appendix A).

Three buoys held a 2-MHz single-point Nortek AquaDopp acoustic current meter recording at a rate of once per 600 s waterflow and tilt (Fig. 1b; 2a). The triangular setup in the same horizontal plane affords the possibility to compute $\zeta$ in (2). While the current-meter data are considerably noisier and less resolved than the T-sensor data, their tilt sensor shows less bias from electronic drift. The nominal instrumental error in horizontal waterflow speed measurements is 0.009 m $s^{-1}$ for each 600-s sample. As the acoustic deep-sea measurements require noise-filtering with cut-off at 4 cpd (cycles per day) per horizontal component, this reduces the precision or one standard-deviation error to $0.009/(216/6)^{1/2}$ = 0.0015 m $s^{-1}$ assuming a Gaussian noise distribution. Consequently, the precision in $\zeta$ amounts about 0.00004 $s^{-1}$. Similarly, the noise-reduced precision in tilt-angle magnitude amounts about 0.004° for current-meter measurements and 0.002° for T-sensor measurements. (T-sensor temperature is a very low-noise measurement, and only requires filter cut-offs at O(1000) cpd.)

### 2.1 General tilt expectations for large-ring mooring array

A vertical-line mooring with a single buoy on top is not displaced exactly like a stiff pendulum, but in a slight J-shape with largest line-tilts near the anchor. In rest with zero waterflow, the upright position is fully due to buoyancy force $F_b$ in negative direction of gravity. For non-zero waterflows that are assumed most horizontal, the drag force $F_d$ by resistance of the mooring line and elements is in disbalance with $F_b$ because they are initially perpendicular to each other from rest (Fig. 2a). A balance of force (amplitudes) is obtained when, to first order [14]–[16],

$$|F_b \tan\alpha| = |F_d| = \tfrac{1}{2} C \rho A |U|^2, \quad (3)$$

where $|U| = (U^2 + V^2)^{1/2}$ denotes the waterflow amplitude, A the surface of flow-obstructing objects, $\rho \approx 1027$ kg m$^{-3}$ the density of seawater and $C \approx 1$ the drag coefficient for a cylinder object. (More about distributed drag on mooring lines can be found in the mooring design software by R. Dewey; https://rkdewey.github.io/Mooring-Design-and-Dynamics/mdd.html, last accessed 14 September 2026.)

For $|F_b| = 1300$ N of a single mooring line [12], $A \approx 125\times0.006 + 1.4\times0.4 = 1.3$ m$^2$ surface of mooring line of height h = 126 m plus buoy, and typical flow speed of $|U| = 0.05$ m s$^{-1}$, (3) provides an expected tilt angle-magnitude of $|\alpha| = 1.7/1300 = 0.07°$. Hereby, $|U|$ is taken constant over the entire mooring line. A more realistic flow includes amplitude and direction changes in the vertical, also over h = 126 m in the deep Mediterranean. A change of $\Delta U = \pm0.01$ m s$^{-1}$ gives $\Delta\alpha \approx \pm0.03°$ around above expected value of $|\alpha|$.

The displaced height of the buoy is then $126(1 - \cos|\alpha|) < 0.001$ m and thus negligible for typical values of waterflow and induced tilt. (Typical horizontal displacements are $126\sin|\alpha| \approx 0.2$ m.) These height displacements are much smaller than the height variations $\Delta h = \pm0.4$ m due to the doming of the large-ring mooring array's cable grid, with largest heights in the centre of the grid [17].

The orientation of the mooring array, pointing towards the north-northwest (Fig. 1b), demonstrates that the three vertical lines equipped with current meters on top form a triangle in the horizontal plain of which the base is closely aligned to the west-east direction. Line 14 is located approximately due north, its perpendicular intersecting the baseline approximately halfway between lines 57 and 35. The doming of the steel-cable grid suggests that line 35 is highest, as it is furthest from the large ring, about 0.5 m above line 14 (lowest) and about 0.2 m above line 57.

## 3 Results

The general overview of three months of stored tilt and other parameter data from h ≈ 125 m above seafloor is presented in Fig. 3. It demonstrates time variability in waterflow which is dominated by mesoscale (~monthly), sub-mesoscale (~1–10 days) and near-inertial variations with time (Fig. 3a). Waterflow speeds seldom exceed 0.06 m $s^{-1}$. The sub-inertial <1-cpd low-pass filtered dominant U-component data from the three current meters demonstrate consistently similar waterflow speeds when |U| ~< 0.03 m $s^{-1}$ (Fig. 3a), near-inertial flow amplitudes are largest (Fig. 3a), temperature is near minimum value of Θ = 12.825°C (Fig. 3b) and when tilt is close to its relative base-value of about 0.07° (Fig. 3c). For comparison, expected values from (3) above base-value amount 0.14±0.03°.

Larger tilt generally associates with larger $|U|^2$ as in (3), but exceptions occur, visibly e.g. on days 330 and 386. Larger tilt generally associates with higher temperatures (Fig. 3b), including on days 330 and 386, and which relates with larger turbulence events [2]. Mostly when |U| >~ 0.05 m $s^{-1}$, the three waterflow speed measurements diverge, with the one from line 14 being often higher than one or both others by up to 0.01 m $s^{-1}$ during the presented 3-month period. (Later episodes without T-tilt measurements showed also higher speeds at other instruments.) As line 14 is north of the other two that are closely aligned to the west-east axis, this may associate with dominant zonal flow (Fig. 3a). The fact that line 14 is about 0.5 m lower than line 35 is not expected to cause great effect, as it would imply a vertical current shear of up to 0.02 $s^{-1}$, or 200 times N ~ f, which is unrealistic.

The array's horizontal waterflow differences result in non-negligible ζ, which adds to the planetary vorticity component f for motions at horizontal scales smaller than 50 m (Fig. 3d). The effective Coriolis frequency, $f_{eff} = (f^2 + f\zeta - \partial U/\partial x \cdot \partial V/\partial y)^{1/2}$ [11], varies by about ±f depending on different zonal-flow conditions. It has values <f when waterflows are generally eastward, and >f when waterflows are westward. Tilt-enhancement of 0.1–0.3° above base-value (Fig. 3c) exceeds relative values expected from (3), and occurs under both zonal-flow $f_{eff}$-conditions. It thus relates with |ζ|, seemingly independent

of the sign of ζ. However, the horizontal difference in tilt (Fig. 3e) changes sign with $f_{eff}$ and thus relates with ζ especially consistently in the west-east direction between lines 57 and 35. As tilt is a magnitude without direction sensitivity, this implies that waterflows to the east with negative relative vorticity induce more tilt on the southwest side of the array compared to its east side, and vice-versa. Considering also the northern line 14, results are more ambiguous. However, it is unclear what precisely causes the tilt differences, as the wakes from the vertical lines are expected to induce relatively weak turbulence due to the small diameters of line and buoy so that vortex shedding is not expected to be noticeable over scales O(0.01) m for the line and O(1) m for the buoy.

Although tilt-angle amplitude associates with temperature variations, no direct quantitative connections can be made with turbulence processes and possibly their intensity. This is because the tilt measurements are limited in resolution of the full turbulence range, capturing only the largest eddies at best (Appendix B; Fig. A2). Most of registrations are resolving the IGW and sub-mesoscale motions governing the turbulence processes. However, the apparent white-noise part of tilt demonstrates qualitative information on turbulence intensity (Fig. A3).

Spectra in Fig. 4 confirm the association of tilt with flow amplitude (and with amplitude squared, not shown), as these show a distinct peak around f, and a particularly steep slope $p < -2$ of power-law $\omega^p$ in the sub-mesoscale frequency range $0.1 < \omega < 0.4$ cpd, which is part of the sub-inertial range $\omega < f$. (This slope extends to the range $0.1 < \omega < 0.8$ cpd for the kinetic energy spectrum, not shown, which has a sharper f-peak than |U| that is slightly blue-shifted from f rather than |U|'s red-shift.) That sub-inertial range is at frequencies $\omega < \omega_{min}(N=f)$ largely below those of the IGW band, for N = f, and considered to largely contain energy from sub-mesoscale eddy motions. However, given the variability in effective inertial frequency, hence relative vorticity, and regularly occurring $f_{eff} \rightarrow 0$ (Fig. 3d), it may include freely propagating small-scale IGW at sub-inertial frequencies as $\omega_{min} \rightarrow 0$.

The best-fit slope for |α|- and |U|-spectra is around $p = -11/5$, which complies with the power-law model of buoyancy-driven motions by [18] and [19], henceforth BO, but errors are considerable. Nonetheless, deviations tend away from other power-law model-slopes like $p = -2$ for finestructure contamination [20], [21] and internal waves established for open seas away from boundaries at $f << \omega$

<< N [22], p = −5/3 established for the inertial subrange of passive-scalar shear-induced turbulence [23], [24], and, not shown, less steep p = −1 for intermittency of a deterministic chaos process [25], [26].

In contrast, the tilt's horizontal difference shows a distinctly different spectrum with a weaker near-inertial peak (Fig. 4). Except for an unknown sub-peak at 0.5 cpd, its spectral slope in the sub-inertial, sub-mesoscale frequency range matches p = −7/5, and is significantly different from that of tilt and kinetic energy. It complies with the BO-model for active scalars, which thus have a different slope from that of the associated advective-flow kinetic energy. However, it matches the sub-mesoscale slope of relative vorticity, and that of its magnitude, and it matches the slope of scalar quantity temperature, which suggests dominant buoyancy-driven motions.

The limited spectral information from only three current meter data records can be extended by exploring variations in tilt and tilt difference between various T-sensors at different lines (Fig. 5). Tilt is spectrally similarly varying vertically at the top and bottom of a given line, which however does differ in variance by a constant value at all frequencies, the larger values at the bottom, as expected for a single top-buoy mooring line forming a J-shape. The spectra for horizontal tilt-difference signals are indeed less featured, and show considerable variation in variance at all frequency ranges, including the sub-mesoscale range. Nevertheless, the mesoscale bulge and the spectral slope remain statistically significant features.

### 3.1 Comparing tilt and vorticity at sub-inertial frequencies

Although to first order mooring-line tilt-angle magnitude is imposed by $|U|^2$ due to the drag force, without considering a particular direction, a tight constant relationship as in (3) is not achieved in practice. This is partially due to the mooring construction that deviates from a stiff pendulum, but probably more due to flow variations at short spatial scales. While vertical flow variations are unknown, because not measured, at the deep Mediterranean site, horizontal variations have been captured in $\zeta$.

The observed small-scale $|\zeta| = O(f)$ are very large compared to reported open-ocean values of <0.2f [11], but they are commensurate with, albeit larger than, mid-depth Mediterranean drifter observations of about 0.5f [1]. Associated with peak-values of $|\zeta| = 4f$ are $|\Delta U| = |\Delta V| = 0.02$ m s$^{-1}$ over horizontal scales of $|\Delta x|$, $|\Delta y|$ = 50 m, i.e. a flow difference of 0.004 m $s^{-1}$ over 10 m, at sub-inertial scales. Using an aspect ratio of 1, this would yield a shear magnitude $|S| = 4f$ that complies with observed small-scale stratification layering so that N = 4f, in marginal stability with gradient Richardson number $Ri = N^2/|S|^2 \approx 1$. As a result, the relative vorticity magnitude potentially reflects vertical flow differences that seriously affect mooring-line tilt-angle magnitude in addition to large-scale flow speeds.

Investigation of a one-month detail in Fig. 6 shows the ambiguity of relationship between $|\alpha|$ and $\zeta$. In the comparison plots here and elsewhere, $\zeta$ is scaled, somewhat arbitrarily, by 2f at the fixed mid-latitude site, as to first approximation $f_{eff}/f = (1 + ½\zeta/f)$ [11]. The relationship between tilt amplitude and relative vorticity changes from out-of-phase between days 320–330 to in-phase between days 340–350. A more consistent relationship is found between $|\alpha|$ and $|\zeta|$, which implies that the sign of relative vorticity does not matter. In general, correlation is good, although fails over short periods like around day 330 when relatively intense turbulence [2] and warm waters are observed (Fig. 3b). In comparison for this monthly period, a positive relationship is found between $|\alpha|$ and $|U|$ for $|\alpha| \sim< 0.13°$ (the dotted line in Fig. 6), and a negative relationship for larger $|\alpha|$.

### 3.2 Details of a variable warming period

A six-day period governed by easterly waterflow, slantwise warming from above/sideways, including relatively large $|\alpha|$ and $|\zeta|$ is investigated in more detail in Figs 7–9. In Fig. 7 time series are shown of waterflow, relative tilt-angle magnitude, negative relative vorticity, and temperature with depth.

When slantwise warming is relatively weak up to day 327, waterflow speeds are often consistent between the three current meters (Fig. 7a), except that the (east-component of the) inertial/IGW motion of line 14 wanders in- and out-of-phase with one of the other two. After day 327 more differences occur between the current meters 50 m apart horizontally with typical values of 0.01–0.02 m $s^{-1}$, also when the total waterflow speed is about 0.02 m $s^{-1}$, at day 330, when general relative mooring-line tilt is largest (Fig. 7b).

Thus, tilt is not only dependent on local waterflow speed, but, occasionally, also on that varying elsewhere. Tilt measured by current meter and upper T-sensor are generally found consistent, also at near-inertial scales. Some deviations occur at super-inertial scales, because the T-sensor has better tilt resolution, and also at sub-inertial scales at which the T-sensor tilt is smaller than the current meter's by 0.01–0.02° when |U| > 0.04 m $s^{-1}$, and larger by about the same amount when |U| < 0.04 m $s^{-1}$.

Only during the first 3 days of the record, sub-mesoscale tilt follows negative relative vorticity, with a >1 day delay towards the end of the six-day period (Fig. 7c). On the other hand, a consistent correspondence exists between temperature time-depth variations (Fig. 7d) and tilt, with largest turbulent warm-water pulses following tilt by a few hours up to one inertial period. This is confirmed by the movies associated with Figs 8 and 9, which also show the small-scale variability between different lines. (Warm-water pulses can also precede tilt magnitude by similar differences in time, as is, e.g., observable in Fig. 3 on day 339. This may reflect different timing in horizontal tilt-magnitude differences and sign of relative vorticity. Both parameters spectrally behave similarly as temperature in the sub-mesoscale range in Fig. 4.)

In Fig. 8, the mooring array is viewed from above in the lower panel and the time-line is visible in the upper panel. It shows relative tilt from 10-cpd noise-filtered T-sensor data from 40 upper-level instruments for the period in Fig. 7. The 72-s movie is accelerated so that 2 hours of real-time appear in 1 s movie-time. The associated movie starts with hardly any tilt above base-value, and very little variation between the lines. Soon some light variations are visible, starting at the southwestern side, and notably appearing near the edges of the large ring at about 50-m horizontal distances. This pattern is followed for three consecutive near-inertial periods before intensifying and shorter-scale varying down to 9.5 m. Most intense values and variations occur within the entire ring-interior, around day 330. The movie ends like it started, although some emphasis is now seen on the eastern side of the ring.

In Fig. 9, a quasi-3D cube presents the T-sensor array in the lower panel, with its colour scheme next to the upper-panel time line. The small panel to the right shows the upper-level flow speed by stick diagram for the three current meters. Waterflow speed is noise filtered at 4 cpd, whilst temperature at 3000 cpd. The associated movie has exactly the same duration and acceleration as that of Fig. 8. Nevertheless it appears must faster and more vigorous, with frequent slashes of warm water from above. If the movie would be made slower (e.g. for half-day episodes in [2]), it could be seen that the warming also comes sideways. The waterflow is consistently to the east, but varies considerably with time and between the current meters. All flickering in temperature reflects turbulent motions, as instrumental noise has been filtered out, partially similar in intensity at all lines, but after more detailed inspection also varying between lines. Most intense is the period between days 330.0 and 331.0, about half an inertial period after the most intense tilt period of Fig. 8.

Fig. 10 shows small-scale variations develop over the 3D large-ring mooring array during a 24-minutes detail period. The movie-time is still 20 times faster than real-time, and reflects the slow but persistent character of deep-sea turbulence (N. Oakey, pers. comm.).

## 4 Discussion

Waterflows are best registered by dedicated instrumentation like current meters. If such rather costly instrumentation is not available, one may rely on proxies from instruments like mooring-line tilt sensor. To first order, tilt-angle magnitude is proportional to water-flow speed squared. Such a proxy works best when the flow is uniform in space and purely horizontal, perpendicular to the direction of buoyancy (-gravity). However, waterflows vary over many scales and are thus not uniform also not in the deep sea, as is demonstrated in this paper.

Like any proxy, the effectiveness of tilt mimicking waterflows depends on the instrumental setup. Under uniform flow, a mooring line with buoyancy evenly distributed over its entire length will behave closely like a stiff pendulum, with identical tilt along the line, whereas a single top-buoy mooring line has larger tilt near the anchor. The latter-case J-shape mooring effect is shown here for tilt-angle magnitudes as small as <0.5°. It is therefore not surprising that tilt is observed only partially corresponding with flow-speed squared, although it remains to be investigated why such correspondence is found for small tilt only and even though direct measurements of vertical shear are missing.

Indirectly, the observed relatively large 50-m horizontal flow differences, having values of up to 100% of the mean flow value, demonstrate the presence of high flow variability at the deep Mediterranean site. These differences, and the associated relative-vorticity magnitude, not only affect tilt, with a good correspondence with tilt variations over scales O(10) m, but also the energy transport to motions at other scales and frequencies, varying from 2D sub-inertial eddies to 3D turbulence, which is important for deep-sea life.

The lack of correspondence between tilt and relative vorticity suggests a reversal of cause and effect. Instead of anticipated relative vorticity widening (for negative values) or shifting to higher frequencies (for positive values) of the IGW band, so that for instance near-inertial motions may be trapped or reflected thereby causing variations in shear and stratification, the actual process may be in reversed order. This is brought about by the observation that tilt relates with relative-vorticity magnitude, and probably shear, and precedes convection from above. Slantwise convection-turbulence plumes may increase local vertical stratification, which becomes in balance with shear magnitude (of horizontal flow) to the level of marginal stability. This is diagnosed by relative-vorticity magnitude and, as a proxy, tilt magnitude, which reflects the arrival of an 'underwater squall' by its small-scale motion variations. Paradoxically, in the slantwise direction stratification may become near-homogeneous N ~ 0.3f, so that the IGW bounds may extend their sub-inertial range thereby explaining some of the observed tilt and tilt-difference spectral peaks around 0.5–0.7 cpd independently from the sign of relative vorticity, indirectly. As the convection turbulence is mainly driven by mesoscale variations, this also indirectly relates tilt with atmospheric disturbances [2].

While the observations have been made in relatively weak flows and weak stratification, it may be worthwhile to test tilt as a proxy for stronger water-flows O(0.1–1) m $s^{-1}$ under more stratified conditions, so that convection turbulence may be less dominant. It would be especially interesting to see the variability in tilt between closely spaced mooring lines, whether these would be limited to O(10) m scales, as presumed for turbulence in the density-stratified vertical [27]. The present observations of weak-flow tilt suggest such a limit under the largest observed convection turbulence with variations O(1) times mean tilt, and which may also play a role in a forest under wind forcing.

## 5 Conclusions

Tightly tensioned, single-buoyancy mooring lines tilt <0.5° under weak <0.07 m $s^{-1}$ flow speeds at a deep site in the Northwestern Mediterranean Sea. The tilt-angle magnitude can be used as a proxy for flow speed squared, but to first order only when tilt is <0.13°. Given up to half-inertial period delay from relative vorticity magnitude, mooring-line tilt seems a reasonably proxy for warm-water convection turbulence slanted from above.

Horizontal, and presumed vertical, flow speed variations O(0.01) m $s^{-1}$ cause tilt-angle magnitude variations O(0.1)° over 50-m scales during weak turbulence and over O(10) m scales during more turbulent episodes, as is demonstrated in short movies.

The observed 50-m scale relative vorticity does not relate with tilt and hence does not affect the effective local vorticity by enlarging the IGW bounds.

The proxy for convection turbulence is also found in very high frequency tilt variations, which appear as quasi-random white noise in spectra. Mooring-line vibrations thus extend across a frequency range from sub-inertial O(0.1−1) cpd to Strouhal O($10^5$) cpd, which may reflect energy transfer in an unknown manner.

**Conflict of Interest**

The author declares no conflict of interest relevant to this study.

**Data availability**

The movies to figures 8−10 can be found in van Haren (2026), "Movies to: Short spatial mooring-tilt variations from deep Mediterranean observations", Mendeley Data, V1, https://doi.org/ 10.17632/9bg7z7gz7p.1. Current meter data are available from van Haren (2025): "Large-ring mooring current meter and CTD data", Mendeley Data, V1, https://doi.org/10.17632/f8kfwcvtdn.1.

**Acknowledgements** Captains and crews of R/V Pelagia are thanked for the very pleasant cooperation. I also thank the team of ROV Holland I for the well-performed underwater mission to recover the instrumentation of the large ring. NIOZ colleagues notably from NMF department are thanked for their indispensable contributions during the long preparatory and construction phases to make the unique sea-operation successful. I highly appreciated working with colleagues within the KM3NeT collaboration.

It has been a pleasure and a privilege. The author acknowledges the financial support of Nederlandse organisatie voor Wetenschappelijk Onderzoek (NWO), the Netherlands.

**Appendix A Determination of relative tilt-angle magnitude**

Upper T-sensor tilt-angle magnitude can be referenced at the three vertical lines holding a current meter at the buoy (Fig. A1), with the notion that the different tilt measurements may slightly differ because the buoy is swiveled and has a shackle-connection with the mooring-line. Also, no exact absolute reference was made after the mounting of the current meters to the buoy, so that a slight non-zero tilt-angle magnitude of about 0.05–0.1° is accepted.

Because of the larger electronic drift in T-sensor tilt compared to current-meter tilt, its data are first referenced to the mean value over a short period in time when the waterflow speed $|U| < 0.01$ m s$^{-1}$, so that the mooring line is expected to be close to vertical. Three short periods from the start of the record are identified, on days 305.25, 316.5 and 317.7038. These tune the T-sensor tilt by about ±0.02°. (Unfortunately, no waterflow measurements were made at other levels than at h = 125.5 m above seafloor. Thus, non-negligible waterflow speeds closer to the seafloor attributing to vertical current shear cannot be ruled out during these short periods.)

In general after fitting a non-zero reference base-value of 0.07° and choosing the day-317 reference, a reasonable comparison is obtained between current-meter and T-sensor tilt at line 14. All excursions up from base-value are identified in both types of instrumentation including near-inertial periodic and sub-mesoscale ones (Fig. A1a), although the amplitude of T-sensor tilt in general exceeds that of current-meter tilt during largest excursions such as around day 330, and vice-versa during smaller excursions such as around day 380. It is not clear what causes these different tilt-angle amplitudes between the instrumentations, but it is likely due to the non-stiff connection, and about 1-m different vertical position between buoy and vertical line, and thus causing a different response upon forcing. The shear may also be larger during largest excursions, affecting the line's tilt more than that of the buoy. Such a different response is also found across horizontal distance of about 50 m, with tilt-angle differences amount ±0.1°, well exceeding noise levels by more than one order of magnitude, and with similar magnitude for the two instrumentation systems, albeit not providing identical variations.

**Appendix B Filters and noise**

Apparent instrumental noise level for T-sensor tilt-angle magnitude is reached at about 10 cpd (Fig. A2), which gives half an order of magnitude larger range than for current-meter tilt (Fig. 4). Noise-filtering thus implies low-pass filtering with presented cut-off at 10 cpd (Fig. A2), and 4 cpd for current-meter tilt. Sub-inertial filtering is achieved by applying a low-pass filter with cut-off close to 1 cpd (Fig. A2), which effectively removes most of inertio-gravity waves and retains daily and longer motions, including at (sub-)mesoscales.

The flatness of the T-sensor tilt spectrum in Fig. A2 for $\omega > 10$ cpd suggests random white noise, or a pure instrumental-determined cause. However, it may also contain a quasi-random response of the instrumentation to environmental changes, as mooring-line tilt is strongly dependent on flow magnitude that causes line-oscillations. This is visualized in Fig. A3, in which the three-month record of tilt is inspected in terms of its very high-1000-cpd-pass filtered data. In Fig. A3a, the full record's standard-deviation is shown, which mimics all of the tilt variations in Fig. 3c. Dissemination of tilt from three episodes varying from calm, around day 320 (Fig. A3b), via moderately turbulent around day 325 (Fig. A3c), to relatively intensely turbulent around day 329 (Fig. A3d), demonstrates a non-continuous amplitude variation with similar intermittency albeit at different levels, for all of these. The 1000–2000-cpd spectra show half–one order of magnitude change in variance, besides the general horizontal flatness, mimicking white noise well within error. Some, insignificant, variations with frequency are observed, such as a common local high around 1600 cpd and a local low around 1200 cpd before gradual roll-off to the Nyquist frequency at $\omega > 2000$ cpd, for all three episodes. The response of the instrumental setup thus qualitatively reflects the convection turbulence at all scales of tilt-angle magnitude.

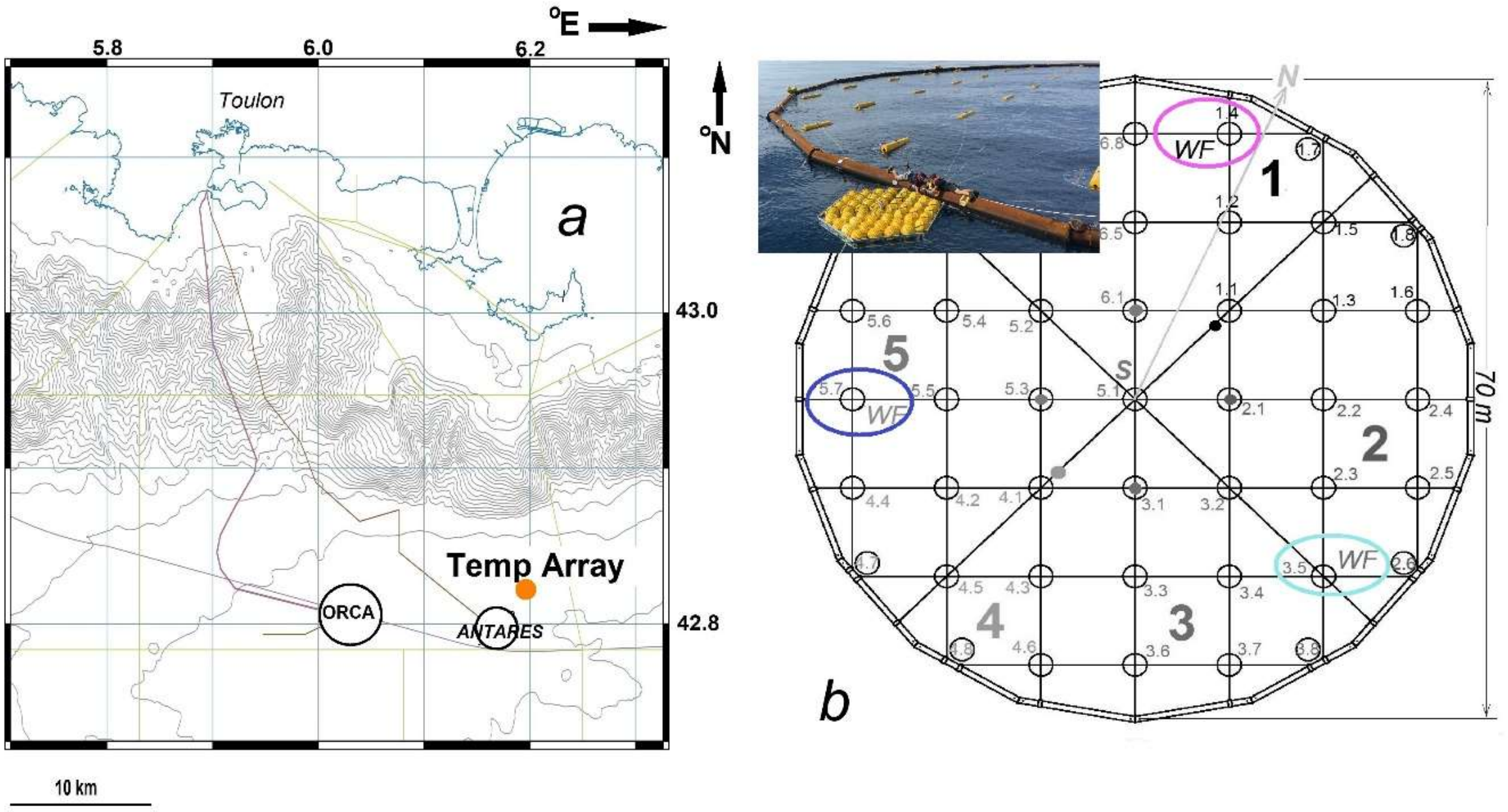

**Figure 1:** Mooring array location and layout. (a) Location named "Temp Array" (orange dot) on map off southern France. The mooring is well east of main neutrino-telescope site “ORCA” of KM3NeT [28] and just northeast of the former “ANTARES” neutrino-telescope site. Isobaths are drawn every 100 m. The grey lines indicate Toulon-harbour approach sectors. (b) Mooring-array orientation at seabed and layout, with steel-cable grid and small rings holding the vertical lines at 9.5 m intervals. Lines are numbered in six synchronization groups. Single synchronizer S is at ring 51. Waterflow ‘WF’ instruments are at buoys on top of lines 14 (indicated in magenta ellipse), 35 (cyan) and 57 (blue). The insert shows part of the large ring just prior to deployment at sea, with free-fall drag-parachute in front.

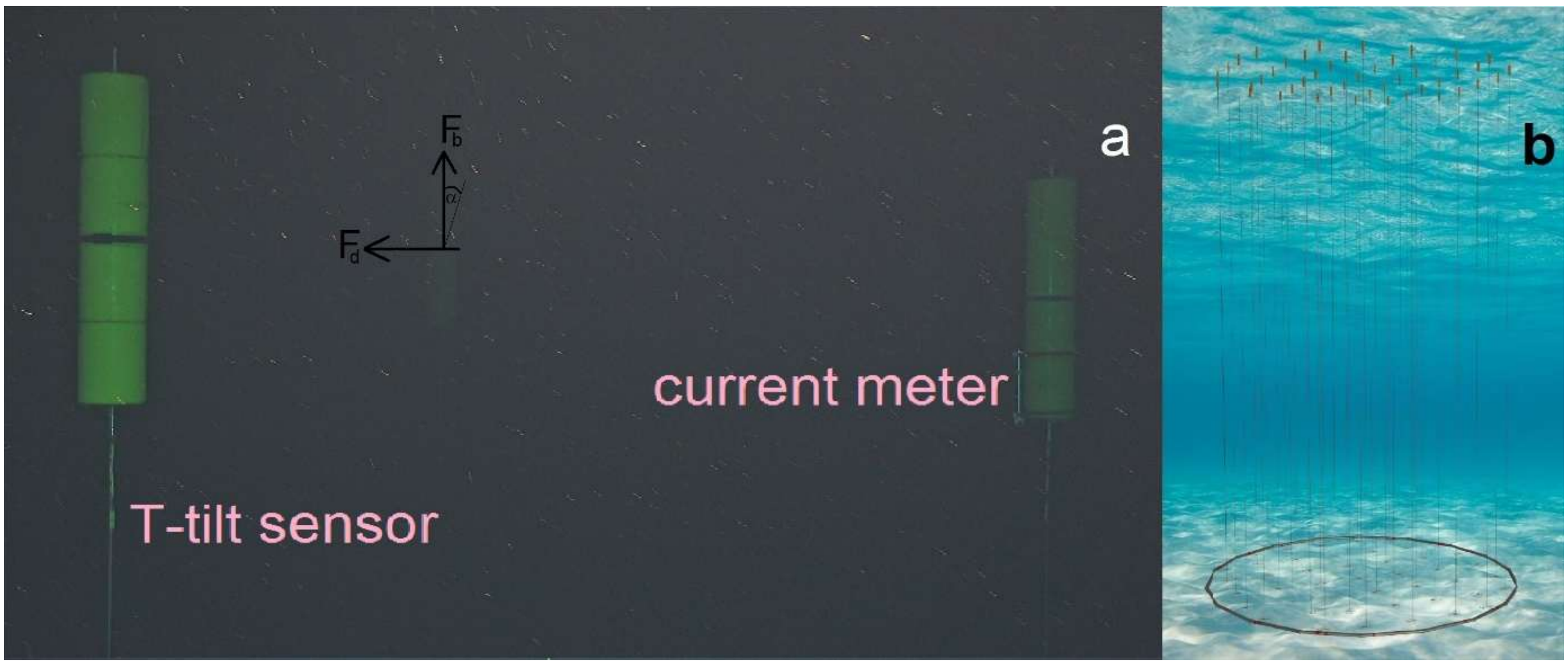

**Figure 2:** Large-ring mooring after unfolding underwater. (a) Three buoys on top of mooring lines. Underneath the left buoy in front, three small instruments are seen taped to the mooring line, up-down: anode for synchronization, uppermost temperature T-only sensor, upper T-tilt sensor. Strapped to the right buoy a downlooking current meter can be seen. Buoyancy and drag forces tilt the mooring line from rest over angle α. Many small suspended creatures and matter are visible as dust-ticks. Video snapshot by ROV Victor (Ifremer, France). (b) Schematic of full array with correct aspect ratio (artist impression by J. van Bennekom).

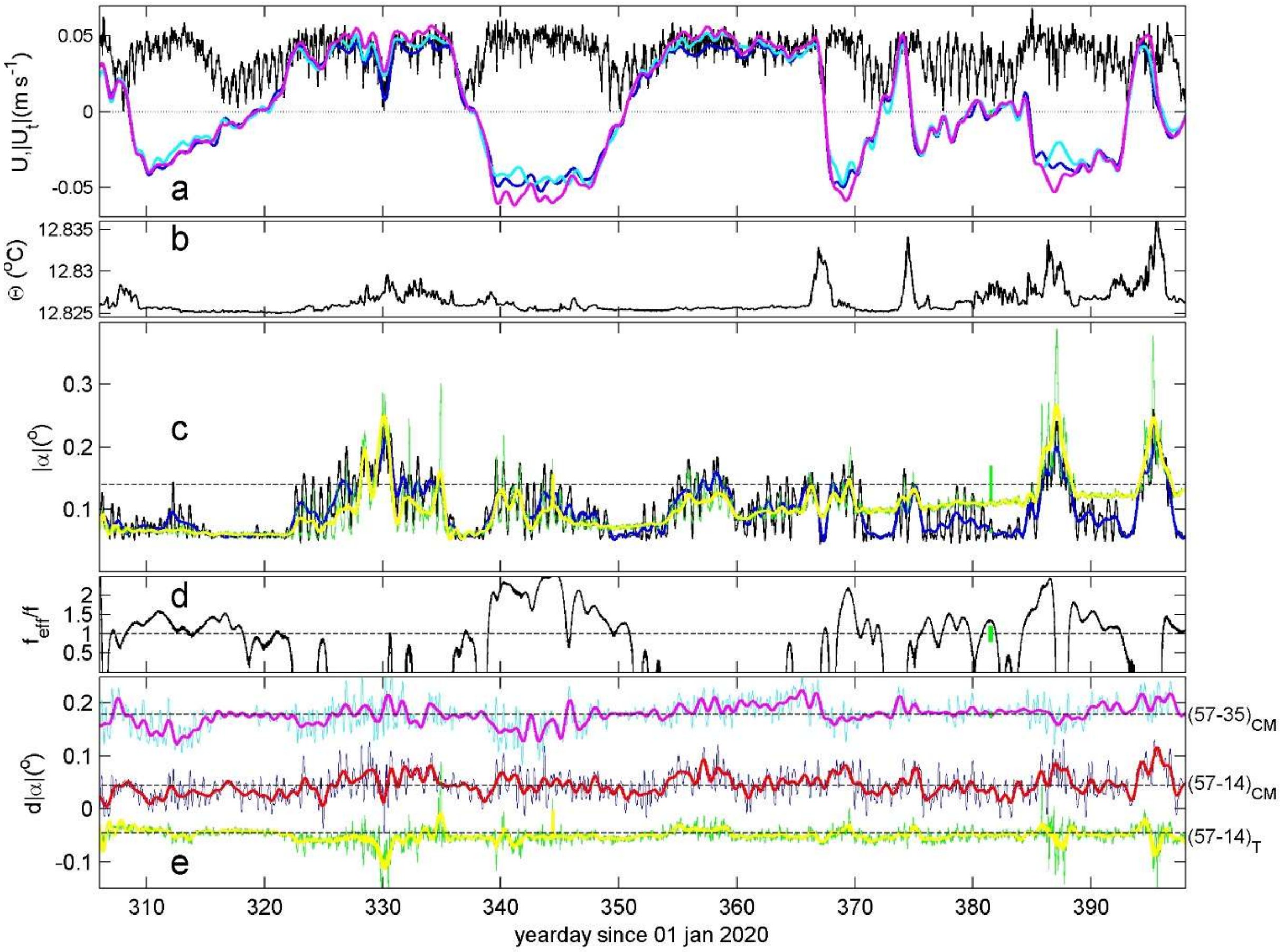

**Figure 3:** Three months timeseries of waterflow and tilt at the top of the mooring array. Error bars are indicated in green on day 381.5. (a) Total noise-filtered waterflow speed (black) measured at line 57, and sub-inertial 1-cpd low-pass filtered East-West component for lines 57 (blue), 35 (cyan) and 14 (magenta). (b) Conservative Temperature [29] measured at upper T-sensor. (c) As a., but for relative tilt-angle amplitude from line 57, in comparison with upper T-sensor tilt data that are not-detrended noise filtered (green) and its sub-inertial data (yellow). The dashed horizontal line indicates the estimate-(3) value above base-value. (d) Effective inertial frequency of combined vertical components of planetary and local relative vorticity for sub-inertial data, scaled with f. (e) As c., but for tilt difference between lines 57 and 14 from T-sensor (green; yellow) and current meter (blue; red), in comparison with current-meter tilt difference between lines 57 and 35 (cyan; magenta). The record pairs are arbitrarily shifted along the y-axis for display purposes.

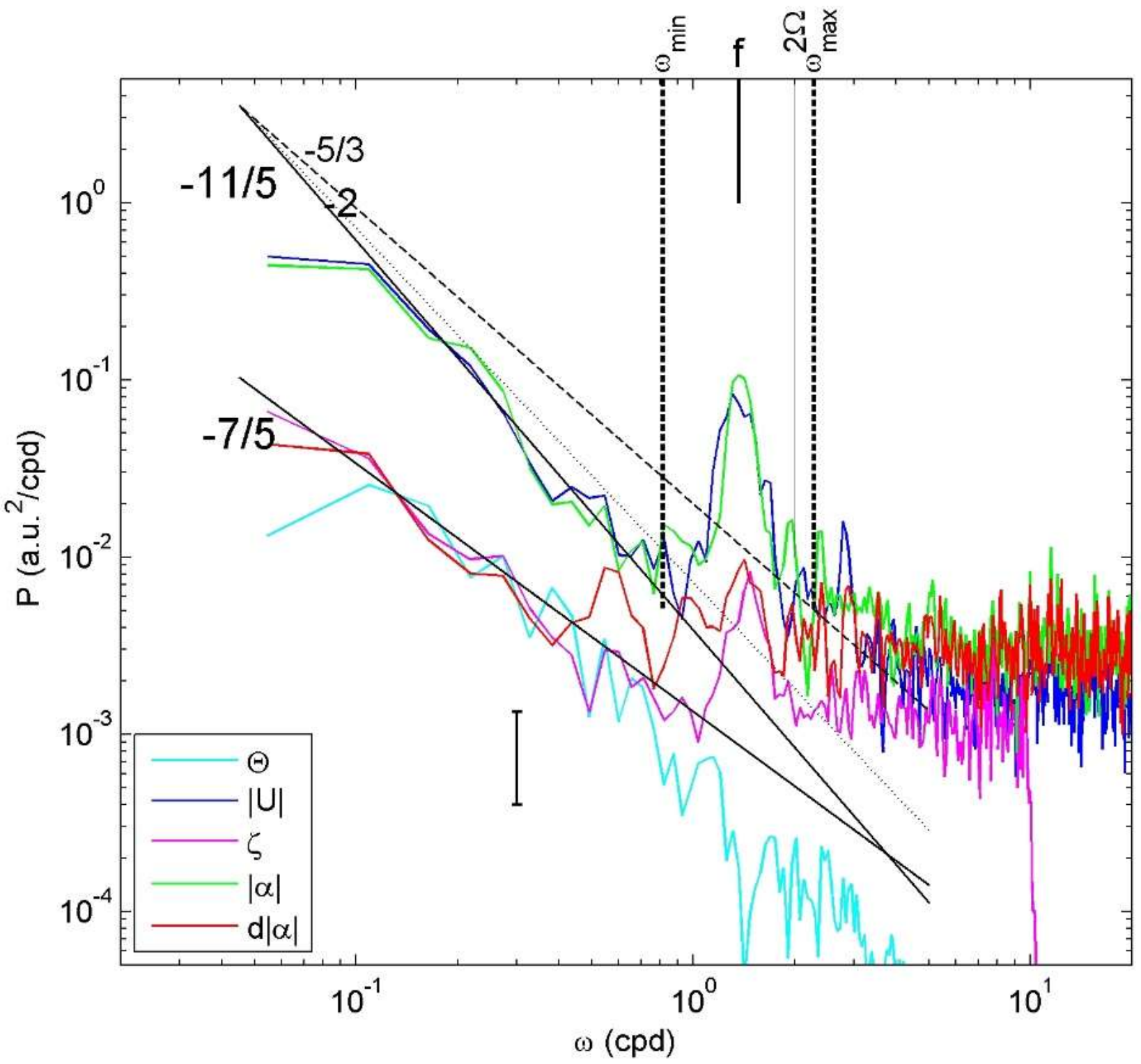

**Figure 4:** Three-month mean power spectra from data at h ≈ 125 m above seafloor for waterflow speed (blue) and 10-cpd low-pass filtered vertical component of relative vorticity (magenta), which are compared with those from current-meter tilt-angle magnitude (green) and its horizontal difference (red; vertically shifted by a factor of 0.45), and with T-sensor temperature (cyan). Indicated are inertio-gravity wave 'IGW' frequency bounds [$\omega_{min}$, $\omega_{max}$](for buoyancy frequency N = f, the inertial frequency, and meridional propagation $\alpha_\varphi = \pi/2$) and semidiurnal frequency 2Ω. Several slopes p of spectral-model power-laws $\omega^p$ are shown (see text). The reader may use a transparent ruler or set square with parallel lines to verify model slopes with observed spectral slopes.

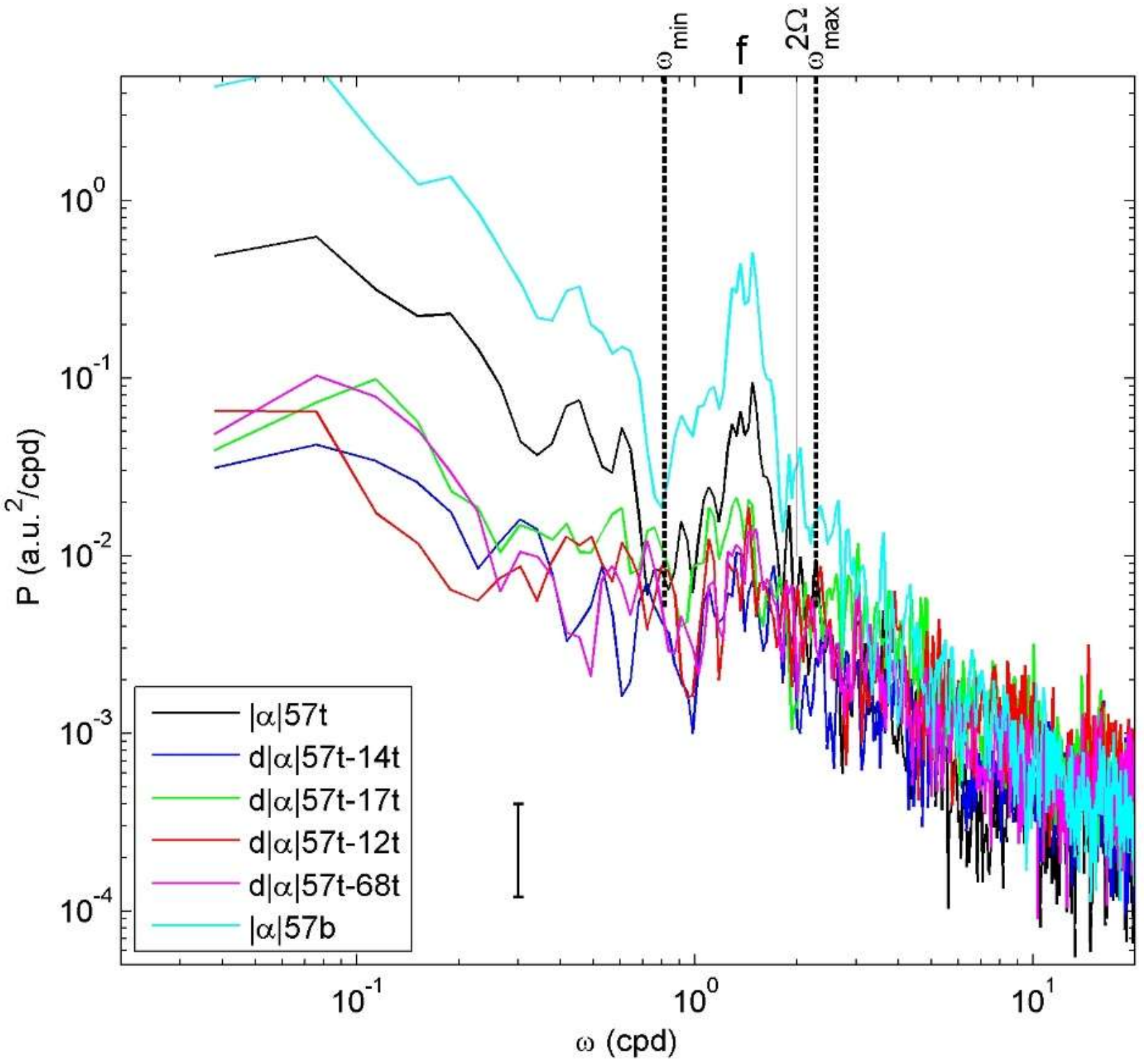

**Figure 5:** Three-month mean power spectra for T-sensor tilt-angle magnitude of line 57 top (black) and bottom (cyan), and top-sensor tilt difference between line 57 and lines 12 (red), 14 (blue), 17 (green) and 68 (magenta). The spectra are a factor of two less smoothed than in Fig. 4, focusing on the entire sub-inertial and IGW bands.

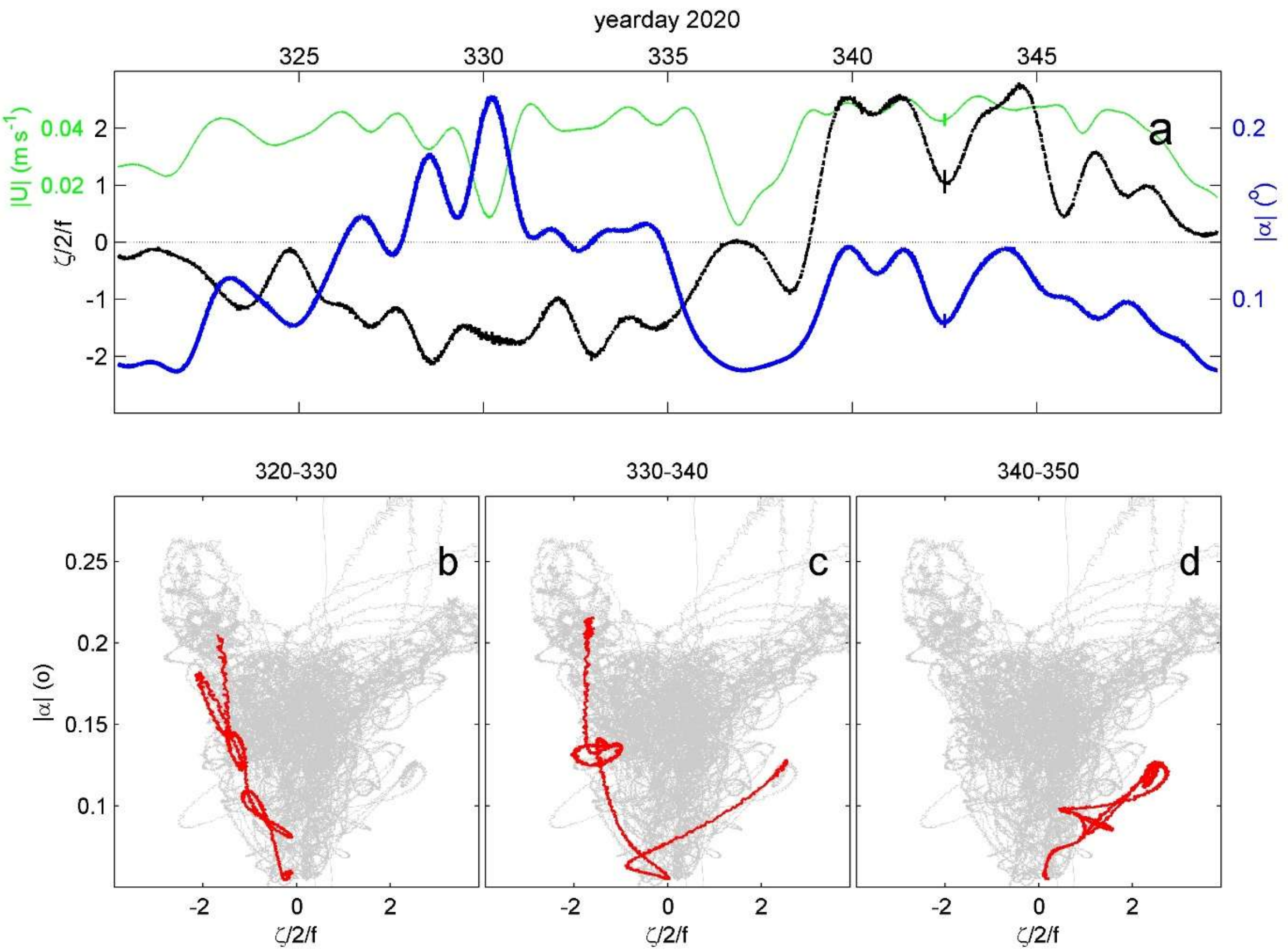

**Figure 6:** Examining the relationship between tilt and relative vorticity. (a) One month time-series with scale of relative tilt-angle magnitude, from current meter at line 57, given to the right. For reference, |U| in green (scale upper left). Error bars are indicated on day 342.5. (b) Relative vorticity scaled with f versus tilt-angle magnitude for first 10 days of a. when average $\zeta < 0$ and $|\alpha|(t) \sim -\zeta(t)$ (dotted red), while average flow $U > 0$ is eastward (Fig. 3a). In background-grey, the entire three months of data. (c) As b., but for middle 10 days. (d) As b., but for last 10 days when average $\zeta > 0$ and $|\alpha|(t) \sim \zeta(t)$, while average $U < 0$ (Fig. 3a).

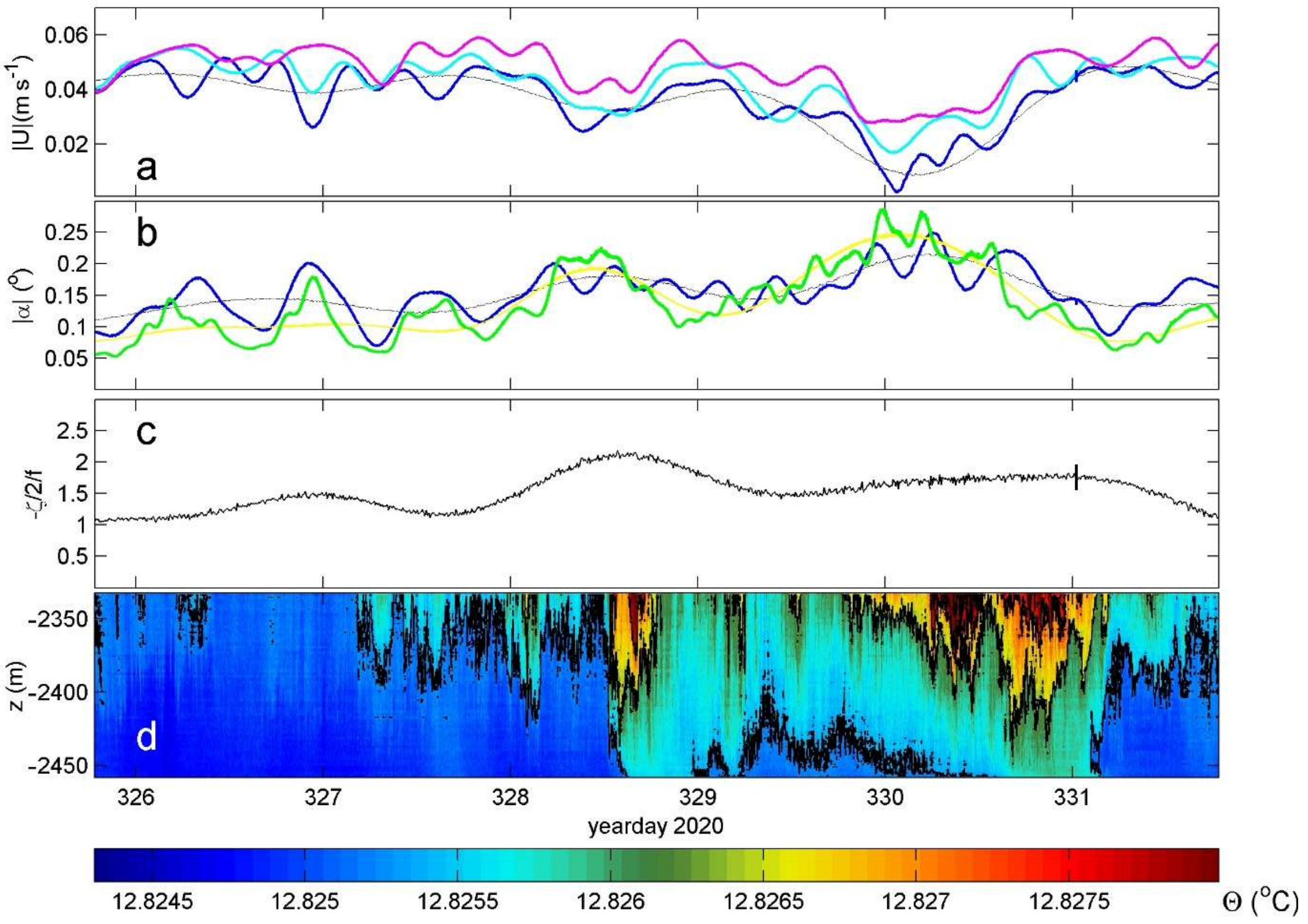

**Figure 7:** Six-day timeseries of waterflow and tilt variations at the top of the mooring array, in comparison with vertical temperature distribution. All data are noise-filtered unless indicated. Error bars are indicated in blue on day 331.03. (a) Waterflow speed from lines 57 (blue; 1-cpd low-pass filtered black), 35 (cyan) and 14 (magenta). (b) Relative tilt from line 57 using current meter data (blue; 1-cpd low-pass filtered black) and using top T-sensor (green; 1-cpd low-pass filtered yellow). (c) Relative sub-inertial vorticity scaled with inertial frequency f. (d) Time-depth series of Conservative Temperature from line 25. Black contours are drawn every 0.001°C.

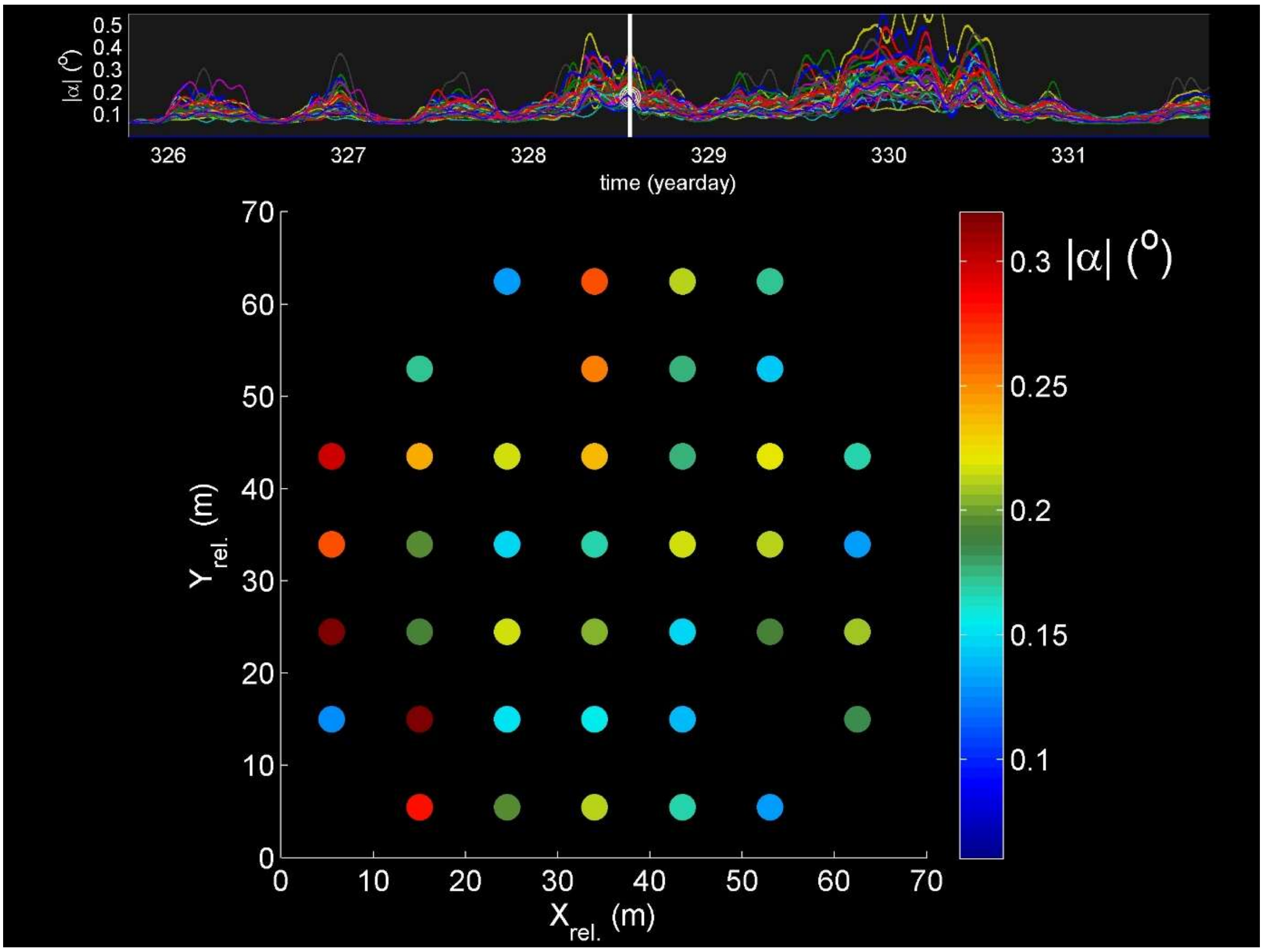

**Figure 8:** Snapshot from 2D movie of 10-cpd low-pass filtered tilt data from 40 upper T-sensors during the period in Fig. 7. In the lower panel the array is viewed from above, and each sensor is represented by a small filled circle of which the colour represents relative tilt magnitude in the scale to the right. In the movie's upper panel, a white time-line progresses in a 6-d timeseries of tilt from all 40 lines. The 72-s movie is accelerated by a factor of 7200 with respect to real-time.

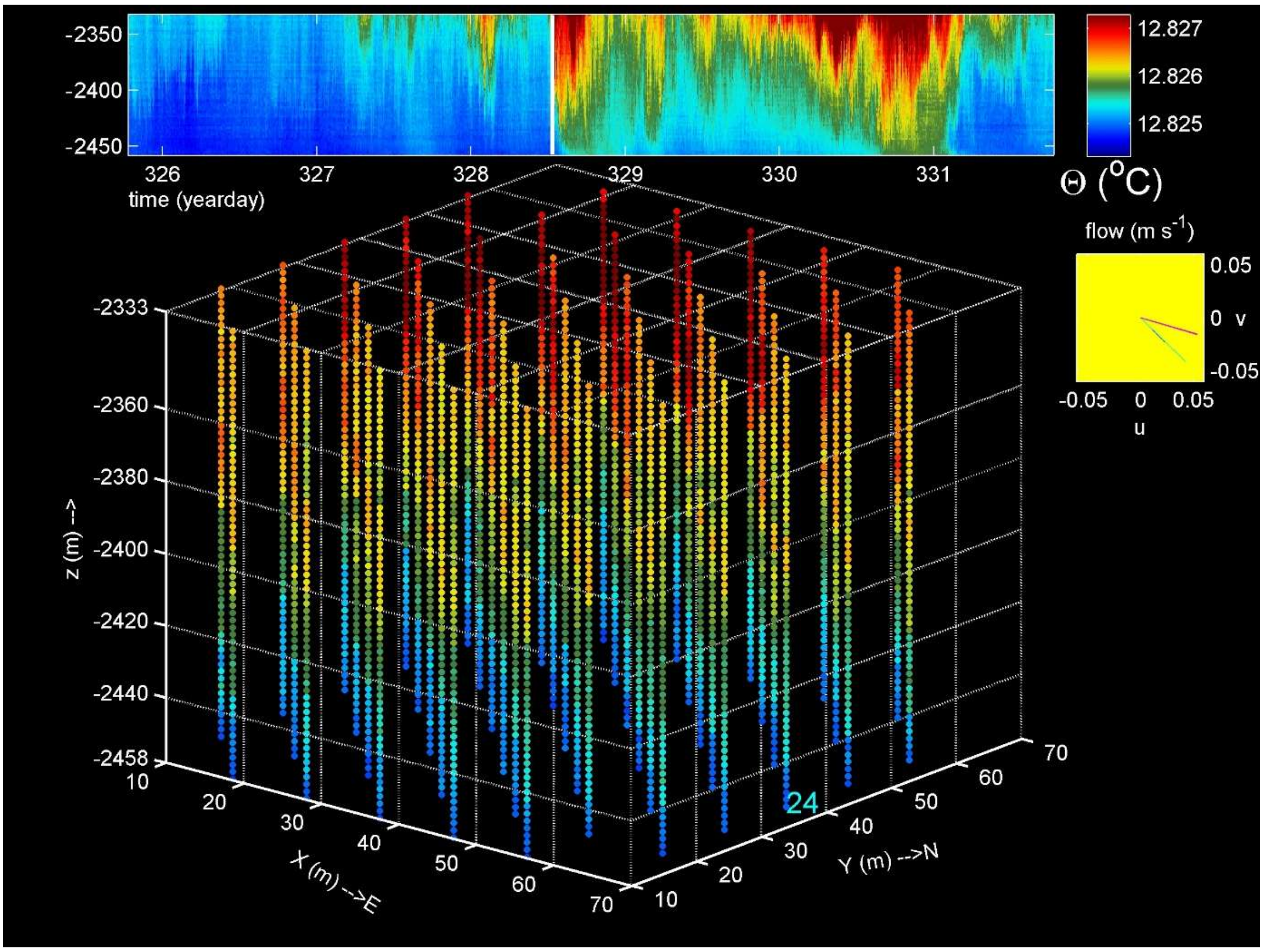

**Figure 9:** Snapshot from quasi-3D movie of 3000-cpd low-pass filtered temperature data from about 2800 T-sensors under stratified water conditions during the period in Fig. 7. In the lower panel's cube, which is vertically depressed by a factor of two relative to horizontal scales, each sensor is represented by a small filled circle of which the colour represents Conservative Temperature in the scale above. In the movie's upper panel, a white time-line progresses in a 6-d/124-m time/depth image from line 24 on the east-side of the cube. To the right of the cube, the noise-filtered waterflow from the three current meters is indicated by vector sticks. The 72-s movie is accelerated by a factor of 7200 with respect to real-time, like Fig. 8.

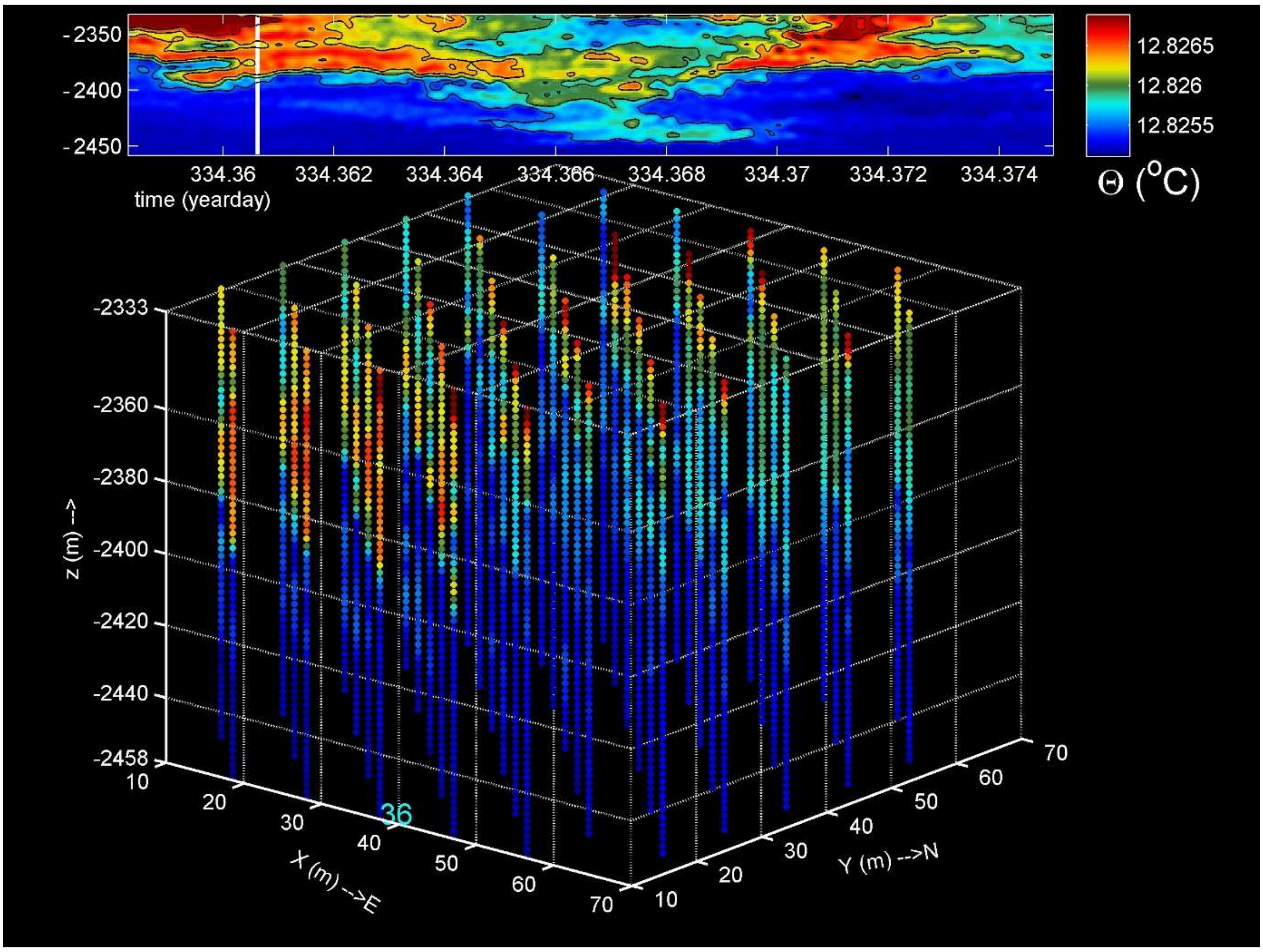

**Figure 10:** Snapshot from a 24-minutes long slow-motion detail movie, which is still 20 times faster than real-time.

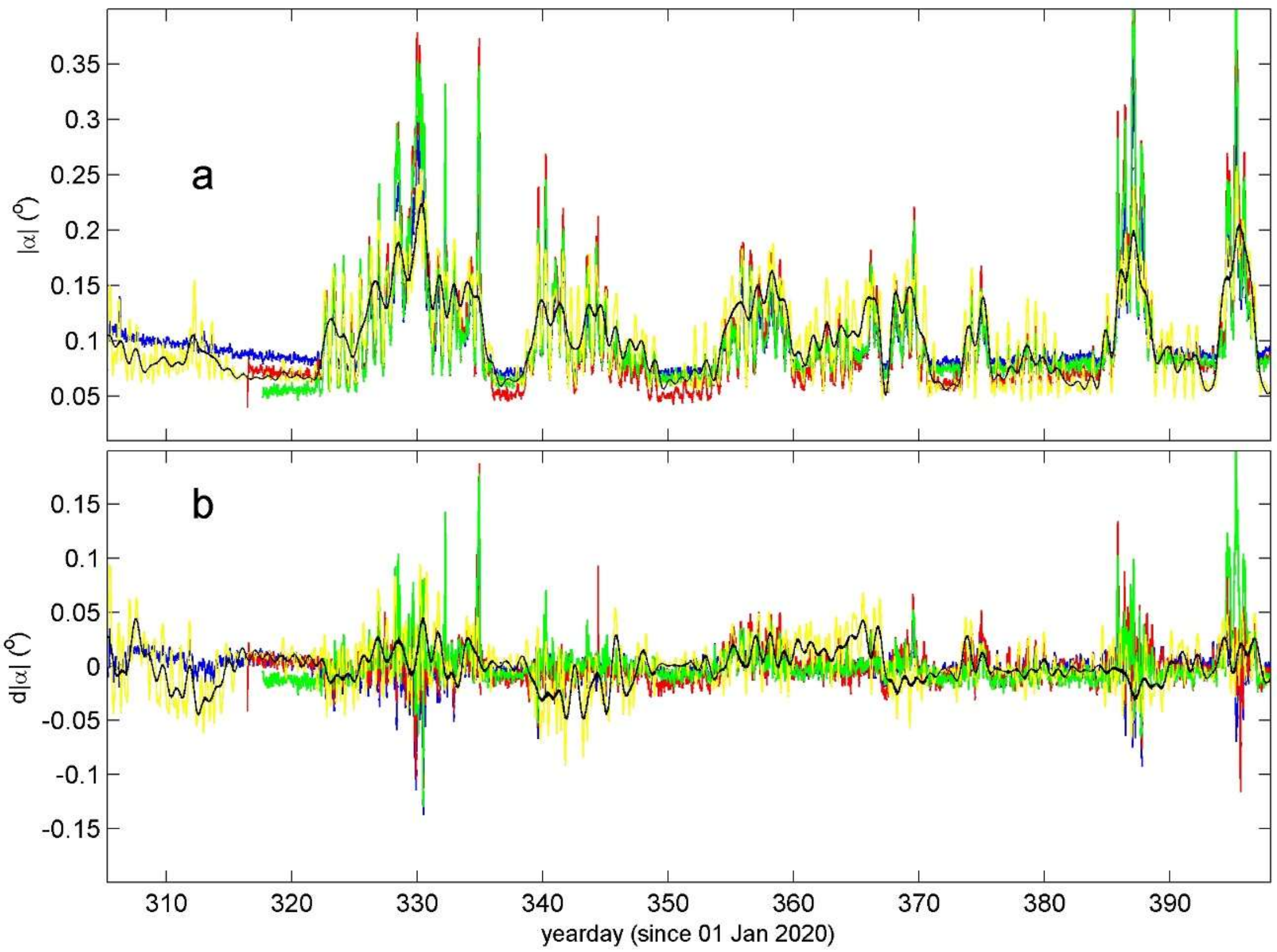

**Figure A1:** Test for reference of line 14 upper T-sensor tilt measurements, noise filtered, as compared with current meter tilt data, noise filtered (yellow) and sub-inertial low-pass filtered (black). T-sensor reference dates with small waterflow < 0.01 m $s^{-1}$ are: days 305.25 (blue record), 316.5 (red) and 317.7038 (green). (a) Detrended noise-filtered tilt-angle amplitude relative to an arbitrary offset. (b) As a., but for detrended tilt difference between lines 57 and 14.

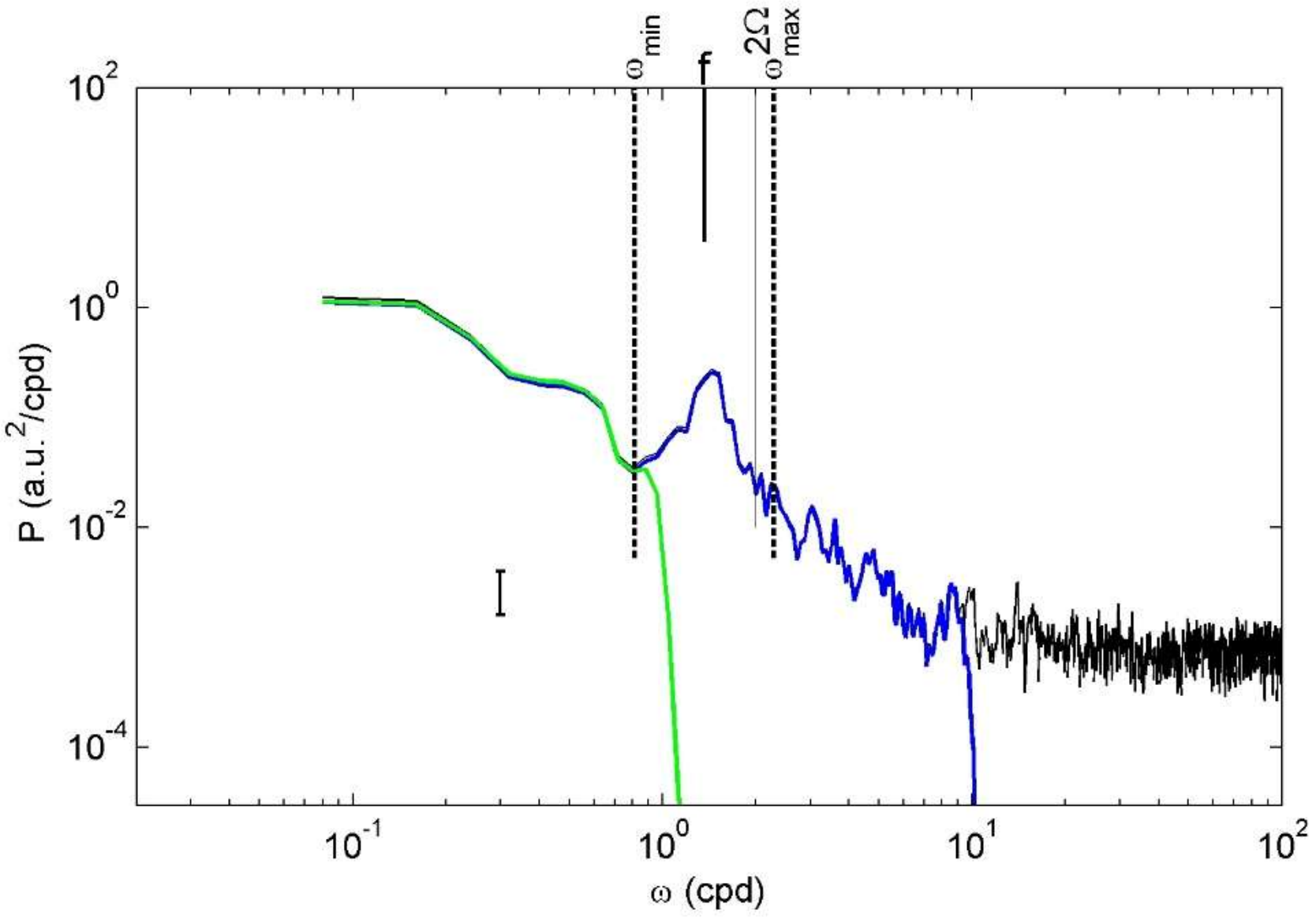

**Figure A2:** Power spectrum showing low-pass filter cutoffs for T-sensor tilt measurements, applied for those of line 57. Original tilt amplitude data (black), noise filtered data with cut-off at 10 cpd (blue), and sub-inertial data with cut-off at 1 cpd (green). (The noise-filter cut-off for current meter tilt is at 4 cpd). The vertical dashed lines indicate the IGW bounds for N = 1.0f and $\alpha_\varphi = \pi/2$, $2\Omega$ indicates the semidiurnal frequency.

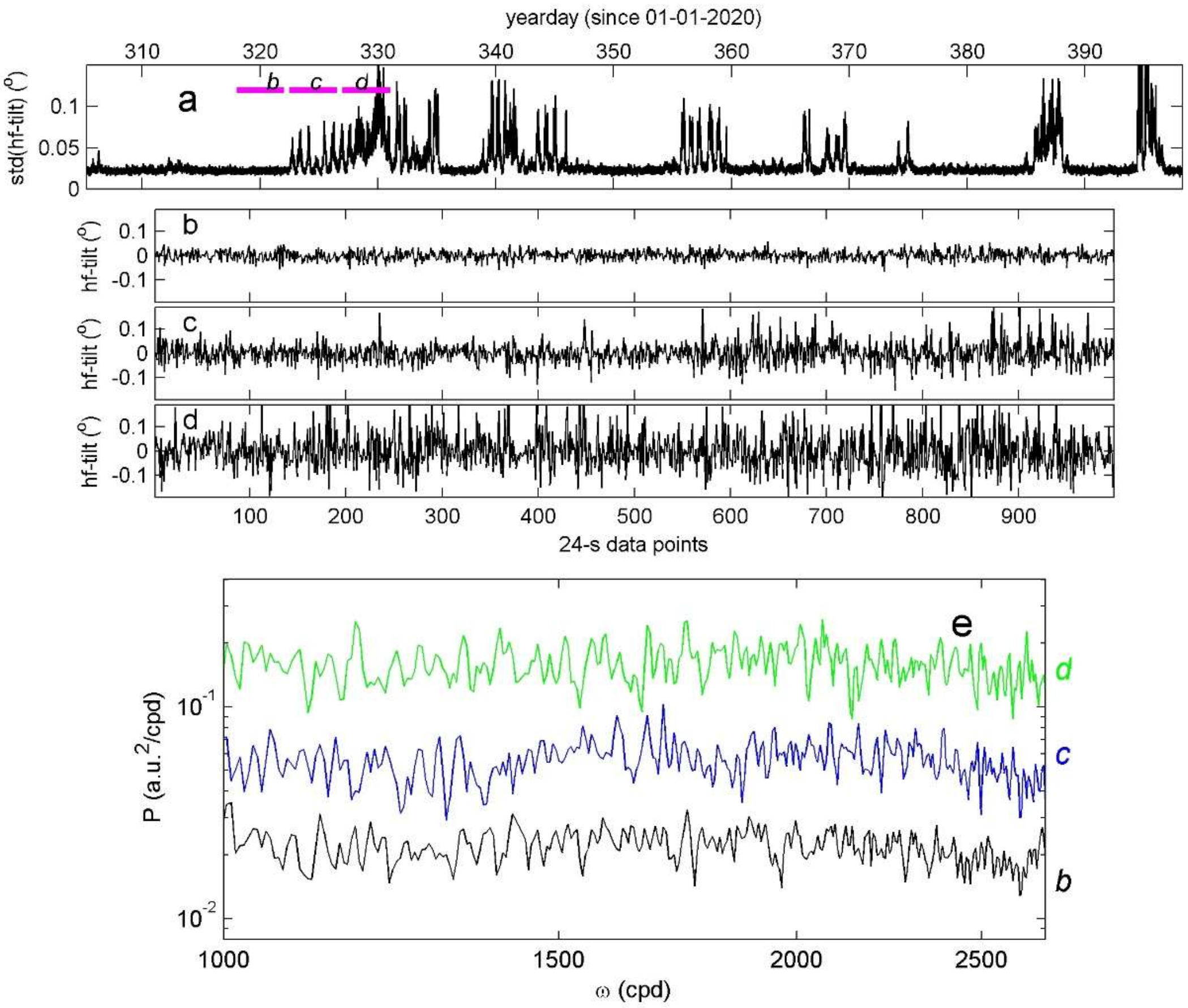

**Figure A3:** "Noise", 1000-cpd high-pass filtered 'hf' data from upper T-sensor tilt measurements of line 17. (a) Three-month time series of standard deviation of noise over 480-s blocks of data. Periods are indicated from which data in panels b.–d. are extracted. (b) Magnification of typical 24000-s time series from period between days 318–322. (c) As b., but for days 322.5–326.5. (d) As b., but for days 327–331. (e) Spectral power from high-frequency parts between 1000 and 2700 cpd of periods in b. (black), c. (blue) and d. (green). The records are not deliberately off-set vertically. Spectra are computed from time series with original (over-)sampling intervals. Application of an anti-alias filter is thus not necessary, because the Nyquist-frequency is more than one order of magnitude higher and its power levels are half an order of magnitude lower than the levels between 1000 and 2000 cpd.